\documentclass[12pt,a4paper,reprint,aps,pra]{revtex4-1}
\usepackage[english]{babel}
\usepackage[T1]{fontenc}
\usepackage[latin1]{inputenc}
\usepackage{microtype}
\usepackage{xspace}
\usepackage{booktabs} %Verbesserte Tabellen (Abstände), midrule, toprule etc.
\usepackage{multirow}
\usepackage{longtable}
\usepackage[tight]{subfigure}
\usepackage{pgfplots}
\pgfplotsset{every axis/.append style={line width=0.7pt},
							legend style = {font={\scriptsize}},
							every axis legend/.append style={nodes={right}},compat=1.3}
\usepackage{calc}
\usepackage{tikz}
\usetikzlibrary{shapes,chains,calc,arrows,positioning}
\usepackage{graphicx}
\usepackage[novolumeabbr]{unitsdef}
\usepackage{textcomp}
\usepackage{amsmath} %Mathematikzeichensatz
\usepackage{lmodern}
\usepackage{pifont} %Sonderzeichen, Pfeil: \ding{225}
\usepackage[pdfpagelabels,plainpages=false,bookmarksnumbered]{hyperref}
\hypersetup{%
	pdftitle={A discrete model for the apparent viscosity of polydisperse suspensions including maximum packing fraction},
	pdfsubject={Submitted},
	pdfauthor={Aaron Dörr, Amsini Sadiki, Amirfarhang Mehdizadeh},	pdfkeywords={multiphase flow, suspension, viscosity, polydispersity, construction process, hard spheres}
}
\usepackage{ltxtable}%muss nach hyperref geladen werden, da sonst Fußnotenlinks nicht funktionieren

\def\gpunkt{\dot{\gamma}\xspace}
\def\dphi{\Delta\phi\xspace}

\def\Vges{V_{T}}

\def\eins{\text{\ding{192}}\xspace}
\def\zwei{\text{\ding{193}}\xspace}

\def\Furnas{{Furnas}\xspace}
\def\Farris{{Farris}\xspace}
\def\Einstein{\mbox{{Einstein}}\xspace}
\def\Bruggeman{\mbox{{Bruggeman}}\xspace}
\def\Roscoe{{Roscoe}\xspace}
\def\Krieger{{Krieger-Dougherty}\xspace}
\def\Mooney{{Mooney}\xspace}

\def\Quemada{{Quemada}\xspace}

\def\Peclet{{Péclet}\xspace}

\def\Brown{{Brown}\xspace}
\def\Boltzmann{{Boltzmann}\xspace}
\def\Stokes{{Stokes}\xspace}
\def\Taylor{{Taylor}\xspace}

\def\McGeary{{McGeary}\xspace}
\def\Sudduth{{Sudduth}\xspace}

\def\Poslinski{{Poslinski}\xspace}

\makeatletter
\newcommand{\Rmnum}[1]{\expandafter\@slowromancap\romannumeral #1@}
\makeatother
\begin{document}

%\includegraphics[bb=0 0 200 200 ]{Externalize/Paper_Part_I-figure0.pdf}
%\fancyheader[EL,OR]{\thepage}
%%%__________________________________________________________________________________
%%%----------------------------------------------------------------------------------
%%%Makros
\newcommand{\etaint}[1]{\left[\eta\right]_{#1}}
\newcommand{\euler}[1]{\mathrm{e}^{#1}}
\newcommand{\kapitel}[1]{Sec.~\ref{#1}\xspace}
\newcommand{\anhang}[1]{Anhang~\ref{#1}\xspace}
\newcommand{\abschnitt}[1]{Sec.~\ref{#1}\xspace}
\newcommand{\seite}[1]{page~\pageref{#1}\xspace}
\newcommand{\bild}[1]{Fig.~\ref{#1}\xspace}
\newcommand{\tabelle}[1]{Tab.~\ref{#1}\xspace}
\newcommand{\gleich}[1]{Eq.~\eqref{#1}\xspace}
\newcommand{\ungleich}[1]{inequality~\eqref{#1}\xspace}
\newcommand{\einsdurch}[1]{\frac{1}{#1}}
\newcommand{\definiert}{\mathrel{\mathop:}=}
\newcommand{\definiertb}{=\mathrel{\mathop:}}
\newcommand{\dtotal}{\mathrm{d}}
\newcommand{\total}[2]{\frac{\dtotal #1}{\dtotal #2}}
\newcommand{\partiell}[2]{\frac{\partial #1}{\partial #2}}
\newcommand{\klammera}[1]{(1-\phi_{#1}^*)}
\newcommand{\klammerb}[1]{\left(1-\frac{\phi_{#1}^*}{\phi_c}
\right)}
\newcommand{\totableitung}[2]{\frac{\dtotal #1}{\dtotal #2}}
\newcommand{\Is}{\Rmnum{1}_S}
\newcommand{\IIs}{\Rmnum{2}_S}
\newcommand{\IIIs}{\Rmnum{3}_S}
\newcommand{\tr}{\mathrm{tr}\,}
\newcommand{\relvis}{relative Viskosität~$\eta_r$\xspace}
\newcommand{\volant}{Volumenanteil~$\phi$\xspace}
\newcommand{\quer}[3]{\overline{#1}_{#2}^{\raisebox{-0.65ex}{\scriptsize$#3$}}}
\newcommand{\marki}[2]{\tikz\color{#1}\pgfuseplotmark{#2};}
\newcommand{\marko}[2]{\tikz{\draw [#1,#2] (0,0) -- (0.08,0);}}

\title{A discrete model for the apparent viscosity of polydisperse suspensions including maximum packing fraction}
\date{\today}
\author{Aaron Dörr}
\affiliation{Institute of Nano- and Microfluidics,  Technische Universität Darmstadt, Alarich-Weiss-Str.~10, 64287 Darmstadt, Germany}
\affiliation{Center of Smart Interfaces, Petersenstr. 32, 64287 Darmstadt, Germany}
\author{Amsini Sadiki}
\affiliation{Institute of Energy and Power Plant Technology, Technische Universität Darmstadt, Jovanka-Bontschits-Str.~2, 64287 Darmstadt, Germany}
\affiliation{Center of Smart Interfaces, Petersenstr. 32, 64287 Darmstadt, Germany}
\author{Amirfarhang Mehdizadeh}
\affiliation{Institute of Energy and Power Plant Technology, Technische Universität Darmstadt, Jovanka-Bontschits-Str.~2, 64287 Darmstadt, Germany}

\onecolumngrid
{\noindent \small \textit{The following article has been published in Journal of Rheology 57 (3), 743-765, 2013. DOI: 10.1122/1.4795746}}
\bigskip
\twocolumngrid

\begin{abstract}

Based on the notion of a construction process consisting of the stepwise addition of particles to the pure fluid, a discrete model for the apparent viscosity as well as for the maximum packing fraction of polydisperse suspensions of spherical, non-colloidal particles is derived. The model connects the approaches by \Bruggeman and \Farris and is valid for large size ratios of consecutive particle classes during the construction process, appearing to be the first model consistently describing polydisperse volume fractions and maximum packing fraction within a single approach. In that context, the consistent inclusion of the maximum packing fraction into effective medium models is discussed. Furthermore, new generalized forms of the well-known \Quemada and \Krieger equations allowing for the choice of a second-order \Taylor coefficient for the volume fraction ($\phi^2$-coefficient), found by asymptotic matching, are proposed. The model for the maximum packing fraction as well as the complete viscosity model are compared to experimental data from the literature showing good agreement. As a result, the new model is shown to replace the empirical \Sudduth model for large diameter ratios. The extension of the model to the case of small size ratios is left for future work.

\end{abstract}

\maketitle
\section{Introduction}\label{chap:Grundlagen}

Multiphase flow is a very important but unexploited field of research according to the variety of unsolved questions. In both nature and technology multiphase flow is rather the rule than the exception. The field of applications includes sprays, bubbly flows, process and environmental engineering, combustion, rheology of blood and suspension systems as well as electro- and magnetorheological fluids. At this point of time, researchers agree neither on the cause of various effects nor on their theoretical description. In this paper, focus is put on suspensions of non-colloidal, non-Brownian hard spheres with a multimodal size distribution. Therefore, the \Peclet number ($\mathrm{Pe}=6\pi\eta_s a^3\gpunkt/(kT)$, where~$\eta_s$ is the fluid viscosity,~$a$ the particle radius,~$\gpunkt$ the shear rate,~$k$ \Boltzmann's constant and~$T$ the temperature), is assumed to be large due to a large particle radius. At the same time, shear rate variations at small shear rate (\cite{1993_Chang,1971_Chong,1992_Shapiro}) result in an apparent viscosity that is independent of shear rate (this limit is confusingly called \emph{high shear limit} by~\cite{1992_Shapiro}). Depending on the length scale of observation as well as on the effects to be described, different approaches may be chosen. On the scale of individual particles in a microscopic approach, besides the strongly restricted possibilities for exact calculation (\cite{1906_Einstein}), {Stokes}ian dynamics (\cite{1989_Russel,1993_Brady}) and Lattice-{Boltzmann} methods (\cite{2005_Hyvaeluoma,2005_Kehrwald}) are employed, for instance. These methods provide a means to investigate the mechanisms occurring in suspensions and to explain the origin of macroscopically observable effects. However, resolving individual particle requests a large computational effort and is therefore not applicable for engineering purposes. Increasing the observation length scale thus leads to reduced computational effort but also to the loss of information because of coarse-graining. The Euler-Lagrange method uses groups of particles---so-called parcels---to represent the particle phase within the fluid carrier phase whereas the Euler-Euler method considers both phases as interacting continuous media (see e.g.~\cite{2005_Chrigui}). Both methods are frequently used to investigate transport, dispersion and reaction processes in dilute and dense suspension systems. In case a pure macroscopic description of the suspension is sufficient, one may model certain flow parameters of the suspension as a whole so that only, for instance, the volume fraction of the particle phase has to be determined during the computation to provide a basis for the calculation of macroscopic suspension properties such as the apparent viscosity. Here, micromechanics models (\cite{2004_Pabst,2002_Torquato}) are well-suited, especially the notion of construction processes employed in the present work. In the effective medium approach, the suspension is constructed from successively added particle size classes, where the newly added particles interact with the present particles as with an effective medium due to large size ratios.\par
In the following, we intend to describe the apparent viscosity as well as the maximum packing fraction of disperse systems by means of the volume fractions of the particle phase, which accounts for the excluded-volume effect governing the apparent viscosity of hard-sphere suspensions (\cite{1994_Wagner}). This restriction of the parameter space by excluding the shear rate corresponds to the asymptotic limit of large \Peclet numbers. Especially, we consider polydisperse suspensions as they are the most general case of dispersions. The apparent viscosity has first been described by \cite{1906_Einstein} in the dilute limit, that is small volume fractions of the particle phase. Later, various attempts to extend the validity of the viscosity relation to higher volume fractions were made, see for instance~\cite{1971_Chong,2004_Pabst,2009_Mendoza,1993a_Sudduth,1993b_Sudduth}.\par
% Taking into account that the apparent viscosity is a macroscopic quantity, individual particles are not considered, rather the particle size is represented by particle size classes characterized by a certain particle diameter. The suspension is assumed to be built up from a finite number of such classes in the course of a construction process that allows for the calculation of the macroscopic suspension properties and thus needs to be described in detail.\par% In a future study, the influence of the shear rate on the apparent viscosity will also be modeled, since suspensions frequently show Non-\Newton{}ian behavior.
The paper is organized as follows. In section~\ref{sec:basics} some basics on apparent viscosity are provided along with an overview of recent work. Section~\ref{chap:viskogleichung} is dedicated to a generalization of the viscosity correlation for monodisperse suspensions. In section~\ref{chap:Grundmodell} a construction process approach used to describe polydisperse suspension viscosity by monodisperse viscosity correlations is presented. Accordingly, a model for the maximum packing fraction of polydisperse systems is developed. The resulting model is then compared to experimental data from the literature. Section~\ref{chap:conclusions} is devoted to conclusions.  

%%#############################################################################################

\section{Basics of apparent viscosity and review of recent works}\label{sec:basics}

%Following the common quasi-single phase formulation we assume the rheological behavior of suspensions to be described by a stress tensor of the form~\cite{2004_Pabst}
%%
%\begin{equation}
%\Tij = \Tij(\rho,T,\Sij)
%\label{eq:allgT}
%\end{equation}
%%
%where~$T$ and $\Sij$ denote the temperature and the symmetric part of the velocity gradient tensor~$\Sij=\left[\nabla\ui + (\nabla\ui)^T\right]$, respectively. We confine ourselves to the case of incompressible flow with a constant deformation history (so-called viscometric flow, see~\cite{2000_Boehme}). Within the framework of continuum mechanics (\cite{1977_Truesdell,2008_Irgens}) these presumptions are met by the constitutive relation of the so-called generalized \Newton{}ian fluid
%%
%\begin{equation}
%\Tij = -p \mathbf{I} +2\eta(T,\gpunkt)\,\Sij
%\label{eq:Newtonallg2}
%\end{equation}
%%
%with the generalized shear rate
%%
%\begin{equation}
%\gpunkt \definiert \sqrt{2\,\tr\Sij^2}
%\label{eq:gpunktallg}
%\end{equation}
%%
%which is equivalent to the second invariant of the symmetric part of the velocity gradient tensor~$\Sij$. The quantity~$\eta$ expresses the apparent viscosity to be determined.\par
Throughout this work, we will disregard the existence of single particles but represent the particles summarily by the so-called \emph{particle size classes}. The presence of particles within the flow increases the viscous dissipation compared with the pure fluid phase with viscosity~$\eta_0$ which leads to the measurability of the apparent viscosity~$\eta_{\text{app}}$. In order to isolate the influence of the particle phase on the apparent viscosity one defines the \emph{relative viscosity}~$\eta_r$ as
\begin{equation}
\eta_r\definiert\frac{\eta_{\text{app}}}{\eta_0}
\label{eq:defetar}
\end{equation}
The apparent viscosity is mainly dependent on the \emph{volume fraction}~$\phi$ 
\begin{equation}
\phi\definiert\frac{V_{\text{particle}}}{V_{\text{fluid}}+V_{\text{particle}}}
\label{eq:defphi}
\end{equation}
of the particle phase, where~$V_{\text{particle}}$ and~$V_{\text{fluid}}$ denote the volumes of the particle and fluid phase, respectively. The volume fraction is sometimes called packing fraction as well. From experiments it is well known that the relative viscosity monotonously increases with increasing volume fraction and exhibits a singular behavior at a value~$\phi<1$. The point where the relative viscosity diverges is commonly denoted as the \emph{maximum packing fraction}~$\varphi_T$ the value of which is not unique but a function of size distribution and flow conditions (\cite{2001_Dames}). For monodisperse systems, where one denotes the specific value of $\varphi_T$ as the monodisperse maximum packing fraction~$\varphi_c$, experiments yield maximum packing fractions of 0.605 (\cite{1961_McGeary,1971_Chong}), 0.55-0.71 (\cite{1992_Shapiro}) or 0.63 (\cite{2001_Dames}). At low shear rate, it is frequently suggested that~$\varphi_c$ can be identified with random close packing,~$\varphi_c\approx0.64$ for spheres (\cite{2004_Pabst}). We will propose a model for the polydisperse maximum packing fraction in \abschnitt{sec:Packungsdichte}.\par
As already mentioned, the fundamental work on the apparent viscosity of disperse systems has been contributed by \cite{1906_Einstein} (erratum~\cite{1911_Einstein}). Therein, the \Stokes equation is solved in a three-dimensional dilatational flow around a spherical particle at rest. Afterwards, the solution is transferred to the case of a suspension with a finite number of particles (volume fraction~$\phi$). The dissipation change due to the presence of the particles leads to the well-known \Einstein relation
\begin{equation}
\eta_r=1+2.5\phi
\label{eq:Einstein}
\end{equation}
\gleich{eq:Einstein} serves a an exact limit for dilute suspensions, that is for~$\phi\to0$. If we denote the~$\phi-$coefficient as \emph{first-order intrinsic viscosity}~$\etaint{1}$ and analogously~$\etaint{m}$ as \emph{$m$th-order intrinsic viscosity}, we can write down the \Taylor series expansion of the relative viscosity, following~\cite{1994_Wagner}, as
\begin{equation}
\eta_r=1+\etaint{1}\phi+\etaint{2}\phi^2+\dotsb
\label{eq:nEntwicklung}
\end{equation}
This representation will be used later in this work. In the literature a great number of viscosity relations of the form
\begin{equation}
\eta_r=\eta_r(\phi,\varphi_T)
\label{eq:H}
\end{equation}
is provided, some of which are listed in Tab.~\ref{tab:Korrelationen}.
\newcommand{\nummer}[1]{\refstepcounter{equation}\quad(\theequation)\label{#1}}
\begin{table}[ht]%
\tabcolsep0.5ex
\centering%
\begin{tabular}{p{10.2em}lr}
\toprule
Reference & Equation & No.\\ 
\midrule
\citet{1959_Krieger} & $\eta_r= \left(1-\frac{\phi}{\varphi_T}\right)^{-\etaint{1}\varphi_T}$&$\nummer{eq:krieger1}\label{test1}$\\
\addlinespace
\citet{1951_Mooney} & $\eta_r=\mathrm{exp}\left(\frac{\etaint{1}\phi}{1-\nicefrac{\phi}{\varphi_T}}\right)$ &$\nummer{eq:Mooney}$\\
\addlinespace
%\textsc{Frankel \& Acrivos} (1967) & \cite{2001_He} & $\eta_r=\frac{9}{8}\frac{\left(\frac{\phi}{\varphi_T}\right)^{\nicefrac{1}{3}}}{1-\left(\frac{\phi}{\varphi_T}\right)^{\nicefrac{1}{3}}}$ &$\nummer{eq:Frankel}$\\
%\addlinespace
\citet{1941_Eilers} & $\eta_r= \left[1+\frac{1}{2}\etaint{1}\left(\frac{\phi}{1-\nicefrac{\phi}{\varphi_T}}\right)\right]^2$&$\nummer{eq:Eilers}$\\
\addlinespace
\citet{1977_Quemada} & $\eta_r=\left(1-\frac{\phi}{\varphi_T}\right)^{-2}$ &$\nummer{eq:Quemada}$\\
\addlinespace
\citet{1949_Robinson} & $\eta_r=1+\etaint{1}\left(\frac{\phi}{1-\nicefrac{\phi}{\varphi_T}}\right)$ &$\nummer{eq:Robinson}$\\
%\textsc{Chong} (1971) & \cite{1971_Chong} & $\eta_r=\left[1+0.75\left(\frac{\phi/\varphi_T}{1-\phi/\varphi_T}\right)\right]^2$ &$\nummer{eq:Chong}$\\
%\addlinespace
\bottomrule
\end{tabular}
\caption[Correlations between relative viscosity and volume fraction]{Correlations between relative viscosity $\eta_r$ and volume fraction~$\phi$ for $0\leq\phi<\varphi_T$}
\label{tab:Korrelationen}
\end{table}
The reason for the existence of such a large number of different correlations is that relations of the form~\eqref{eq:H} do not cover the entire parameter space governing the physical problem. \cite{2004_Pabst} expressed this by the formulation
\begin{equation}
\eta_r=\eta_r(\phi,\,\text{all other details of microstructure}).
\label{eq:etarpabst}
\end{equation}
Clearly the parameter space must be confined to allow for useful modeling and thus the range of validity has to be confined a priori. In the present context, the viscosity relations~\eqref{eq:etarpabst} can be classified in two groups:
\begin{enumerate}
	\item Series expansions with respect to the volume fraction for $\phi\ll1$ according to \gleich{eq:nEntwicklung}
	\item Correlations for~$\phi\to\varphi_T$
\end{enumerate}
Regarding the first group, the \Einstein relation~\eqref{eq:Einstein} with the intrinsic viscosity~$\etaint{1}=2.5$ is commonly accepted as the first order series expansion of the relative viscosity~$\eta_r$ of suspensions with spherical particles, cf.~\cite{2004_Pabst}. However, there is no unique value of~$\etaint{2}$ because of a strong case-sensitivity of this parameter. In \tabelle{tab:etaint2} some examples taken from the literature are listed.
%
%\onecolumngrid
\begin{table}[ht]%
\centering%
\begin{tabularx}{\columnwidth}{@{\extracolsep{1ex}}l@{\extracolsep{1.5ex}}X r@{\extracolsep{0ex}}@{.}l}
\toprule
 Reference & Annotations & \multicolumn{2}{l}{$\etaint{2}$} \\
\midrule
 \citet{1972_Batchelor} & \Brown{}ian motion\newline neglected, random spatial particle distribution & 5&2   \\
 \addlinespace
\citet{1977_Batchelor} & \Brown{}ian motion\newline included, random spatial particle distribution & 6&17  \\
\addlinespace
\citet{1977_Bedeaux} & formalism in wave\newline number space & 4&8  \\
\addlinespace
\citet{1991_Cichocki} & \Brown{}ian motion\newline neglected, {Smoluchowski} equations & 5&00  \\
\addlinespace
\citet{1991_Cichocki} & 
\Brown{}ian motion\newline included, {Smoluchowski} equations & 5&91  \\
\addlinespace
\citet{1959_Krieger} & second-order {Taylor}\newline coefficient of the \Krieger relation~\eqref{eq:krieger1} for $\varphi_T=0.64$, empirical & 5&08  \\
\addlinespace
\citet{1951_Mooney} & second-order {Taylor}\newline coefficient of the \Mooney relation~\eqref{eq:Mooney} for $\varphi_T=0.64$, empirical & 7&03  \\
\bottomrule
\end{tabularx}
\caption[Values of the second-order intrinsic viscosity taken from the literature]{Values of the second-order intrinsic viscosity~$\etaint{2}$ taken from the literature}
\label{tab:etaint2}
\end{table}
%\twocolumngrid
%
The value~$\etaint{2}=5.2$ according to~\cite{1972_Batchelor} will be used in this work because the focus lies on hard-sphere suspensions with purely hydrodynamic interactions.\par
The correlations listed in \tabelle{tab:Korrelationen} are intended to be valid especially for large values of~$\phi$. They all coincide with respect to a singular behavior at the point~$\phi=\varphi_T$, that is when the maximum packing fraction is reached. Note that \gleich{eq:Quemada} does not reduce to the \Einstein relation as~$\phi\to0$. In the next section, an attempt to generalize a viscosity correlation for monodisperse suspensions is presented.

%%#############################################################################################

\section{Generalization of the viscosity correlation for monodisperse suspensions}\label{chap:viskogleichung}
As shown in the previous section, there are two main types of viscosity correlations, namely polynomial and closed correlations. Polynomial correlations are well suited for describing the low-concentration range but do not show divergence for~$\phi\to\varphi_T$. Closed correlations diverge for~$\phi\to\varphi_T$, but cannot show proper asymptotic behavior for~$\phi\to0$ because the second-order \Taylor series expansion is determined a priori through the viscosity relation.\par
In order to combine the low-concentration behavior of polynomial correlations with the high-concentration behavior of closed viscosity correlations, similar to the approach of~\cite{2005_Krishnamurthy}, we derive a heuristic correlation that allows for choosing all of the relevant parameters~$\etaint{1}$,~$\etaint{2}$ and~$\varphi_T$. This can be achieved by considering the two kinds of viscosity correlations as asymptotic limits for small and large~$\phi$, respectively, and matching them. The closed correlation representing the asymptotic large-$\phi$ behavior can be chosen to accurately fit to the experimental data at hand. Here, we choose the \Quemada \gleich{eq:Quemada} (see \abschnitt{sec:Vergleich}), but we also provide the modified form of the important \Krieger \gleich{eq:krieger1}. The method of additive composition (cf. \cite{1975_van_Dyke}) consists of finding the small-$\phi$ behavior of \gleich{eq:Quemada} by \Taylor series expansion up to second order, of subtracting the resulting expansion from the \Quemada expression and of adding the desired second-order expansion~$\eta_r\sim1+\etaint{1}\phi+\etaint{2}\phi^2$ to the result. We end up with the modified \Quemada equation
\begin{multline}
\eta_r(\phi,\varphi_T)=\left(\etaint{1}-\frac{2}{\varphi_T}\right)\phi\\+\left(\etaint{2}-\frac{3}{(\varphi_T)^2}\right)\phi^2+\left(1-\frac{\phi}{\varphi_T}\right)^{-2}
\label{eq:viskogleichung}
\end{multline}
Obviously, the matching only corrects the first- and second-order terms in~$\phi$ as has been expected. As a second example, we write down the modified \Krieger equation
\begin{multline}
\eta_r(\phi,\varphi_T)=\left(\etaint{2}-\etaint{1}\frac{1+\etaint{1}\varphi_T}{2\varphi_T}\right)\phi^2\\+\left(1-\frac{\phi}{\varphi_T}\right)^{-\etaint{1}\varphi_c}
\label{eq:modkrieger}
\end{multline}
where only the second-order term has been corrected because \gleich{eq:krieger1} reduces to the \Einstein relation for~$\phi\to0$.
%
%%#############################################################################################

\section{Development of a viscosity correlation for polydisperse suspensions}\label{chap:Grundmodell}

\begin{figure}[ht]%
{%\renewcommand{\baselinestretch}{1.0}
\centering%
\begin{tikzpicture}[%
rand/.style={fill=red!20,rounded corners,text centered},
pack/.style={fill=teal!20,,text width=3cm,text centered},
normal/.style={fill=blue!20,text width=3cm,text badly centered},
normal2/.style={fill=blue!20,text width=2.1cm,text badly centered},
kreis/.style={fill=blue!20,shape=circle,minimum size=5mm,},
frage/.style={fill=blue!20,text width=3cm,text centered,shape aspect=2,diamond,inner sep=-0.15cm},
start chain=haupt going below,every join/.style={-stealth'},node distance=0.62cm and 0.6cm,every node/.style={minimum width=9mm},line width=0.4pt]
\node (anf) [rand,draw,on chain] {Start $i=1$};
\node (a) [normal,draw,on chain,join=with anf.south] {\linespread{1.0}\selectfont$\dphi_k^i$ from Eq.~\eqref{eq:dphikhochi3} for $k=1,\dotsc,i$\par};
\node (b) [frage,draw,on chain,join,label=175:No,label=-93:Yes] {\linespread{1.0}\selectfont Viscosity\\ relation dependent on $\varphi^{i}_{Ti}$\par};
%Viskosität mit phiT
\node (c) [pack,draw,on chain=haupt,join=with b.south] {$\varphi_{T1}^i=\varphi_c$};
\begin{scope}[node distance=0.3cm][start branch=packung]
\node (d) [pack,draw,on chain,join] {\linespread{1.0}\selectfont $\varphi_{T2}^i$ from Eq.~\eqref{eq:phiTiplus11} or~\eqref{eq:phiTiplus12}\par};
\node (e) [pack,draw,,on chain,join] {$\dotso$};
\node (f) [pack,draw,on chain,join] {\linespread{1.0}\selectfont$\varphi_{Ti}^i$ from Eq.~\eqref{eq:phiTiplus11} or~\eqref{eq:phiTiplus12}\par};
\end{scope}
\node (g) [normal,draw,join,on chain] {\linespread{1.0}\selectfont$\eta_{i}$ from\\ Eq.~\eqref{eq:allgvisko}\par};
%Viskosität einzeln
\node (b2) [normal2,draw,left=of g.west] {\linespread{1.0}\selectfont$\eta_{i}$ from\\ Eq.~\eqref{eq:allgetasterniplus1}\par};
\draw [-stealth'] ($ (b.west) $) -| ($(b2.north)$);
%Zusammenführung
\node (kombi) [kreis,draw,on chain,join=with g.south] {};
\draw [-stealth'] ($ (b2.south) $) |- ($(kombi.west)$);
%Große Schleife
\node (frage) [frage,draw,on chain,join,label=175:No,label=-93:Yes] {$i=n$};
\node (ende) [rand,draw,below=of frage.south] {End};
\draw [-stealth'] ($(frage.south)$) -- ($(ende.north)$);
\node (inkre) [normal,text width=1.4cm,draw,on chain=going left,join] {$i\to i+1$};%war 1.8
\draw [-stealth'] ($(inkre.west) $) |- ($(inkre.west) + (-1.4cm,0cm)$) |- (a.west); %war 0.5cm
\end{tikzpicture}%
\caption[Scheme for the calculation of maximum packing fraction and relative viscosity]{Scheme for the calculation of maximum packing fraction and relative viscosity after the $i$th construction step for~$n$ particle size classes with references to the respective equations}%
\label{fig:Ablauf}%
}
\end{figure}
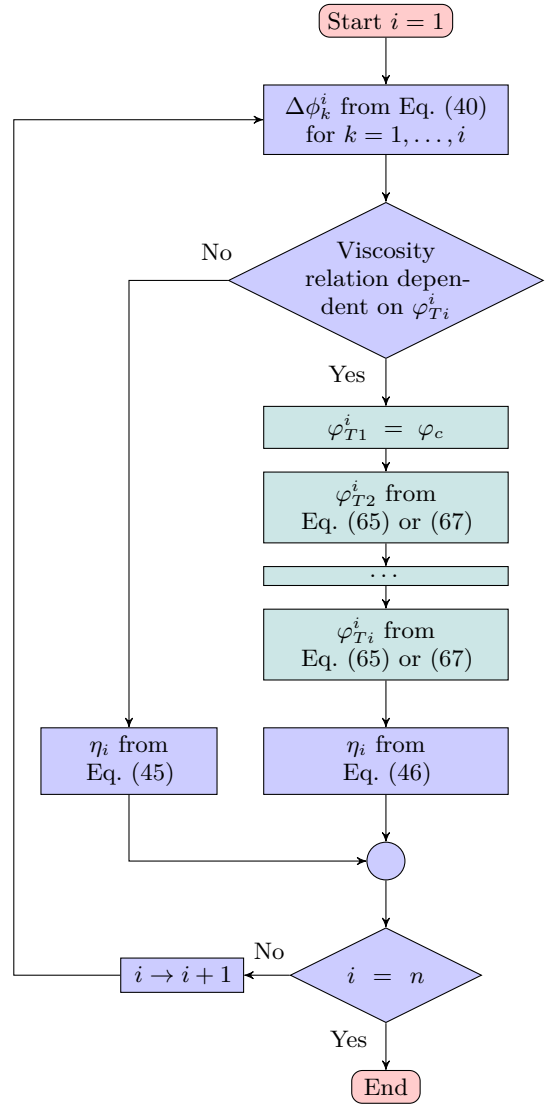

In this section we develop a polydisperse viscosity model based on the notion of a construction process. This approach is first exactly described in \abschnitt{sec:Aufbauprozess}. Subsequently in \abschnitt{sec:Diskretes_Modell} the construction process is applied to the determination of relative viscosity. Then, a model for the maximum packing fraction of polydisperse suspensions is developed in \abschnitt{sec:Packungsdichte} to complete the viscosity model. A graphical scheme provided in Fig.~\ref{fig:Ablauf} may serve for the reader's guidance during the calculation.

\subsection{Starting point: The differential Bruggeman model}\label{sec:Bruggeman}

The differential \Bruggeman model (see also \citet{2009_Hsueh} and more detailed \citet{2002_Torquato}) makes it possible to derive a closed viscosity relation for the full concentration range starting from the \Einstein relation. The \Bruggeman model is also known as \emph{Differential Effective Medium approach (DEM)}. A generalization of the DEM approach is presented by~\citet{1985_Norris}. The \Bruggeman model is based on the notion that an infinitesimal volume fraction of particles is added to an existing suspension with effective viscosity~$\eta_{\text{app}}$ and volume fraction~$\phi$. In the course of this addition it is assumed that the existing suspension can be treated as a homogeneous medium. This can only be valid if the newly added particles have a large diameter compared with the particles already present in the suspension (\citet{1993_Chang}).\par 
We now ask for the change in effective viscosity due to the infinitesimal volume fraction~$\phi^*$ of the newly added particles in the resulting suspension. It can be shown, according to~\citet{2009_Hsueh}, that 
\begin{equation}
\phi^* = \frac{\dtotal\phi}{1-\phi}
\label{eq:diffphistern}
\end{equation}
Note that, in this paper, the total suspension volume is kept constant (cf. \abschnitt{sec:Aufbauprozess}), whereas in~\citet{2009_Hsueh} the volume is variable, leading to a different form of the left-hand side of \gleich{eq:diffphistern}. Because of the small size of the volume fraction~$\phi^*$ we may use the \Einstein relation to describe the change in effective viscosity by
\begin{equation}
\eta_{\text{app}}+\dtotal\eta_{\text{app}} = \eta_{\text{app}}\left(1+\etaint{1}\phi^*\right)
\label{eq:etaplusdeta}
\end{equation}
By inserting \gleich{eq:diffphistern} into relation~\eqref{eq:etaplusdeta} we obtain
\begin{equation}
\eta_{\text{app}}+\dtotal\eta_{\text{app}} = \eta_{\text{app}}\left(1+\etaint{1}\frac{\dtotal\phi}{1-\phi}\right) 
\label{eq:etaplusdetaeingesetzt}
\end{equation}
Integration of \gleich{eq:etaplusdetaeingesetzt} under the initial condition $\eta_{\text{app}}(\phi=0) = \eta_0$ yields
\begin{equation}
\eta_{\text{app}} = \eta_0 \left(1-\phi\right)^{-\etaint{1}}
\label{eq:roscoeohnephic}
\end{equation}
Eq.~\eqref{eq:roscoeohnephic} is known as \Roscoe equation. Since the \Bruggeman model requires a large diameter ratio of consecutively added particle classes, the suspension must consist of a solid phase that can be divided into particle size classes of large diameter ratios. This structure is called hierarchical, see also~\citet{1985_Norris} and~\citet{2002_Torquato}.\par
Another important assumption of the differential \Bruggeman model is the validity of \gleich{eq:etaplusdeta}. The volume fraction of newly added spheres in \gleich{eq:diffphistern} has to be small enough for the \Einstein relation to be valid. This assumes a multiplicative influence of the new particles corresponds to a separation-of-contributions method in contrast to hard-sphere scaling approaches (classification by~\citet{2003_Quin}).\par
We note that the volume fractions~$\phi^*$ may be finite in principle. However, by introducing the infinitesimal increment~$\dtotal\eta$ into the differential \Bruggeman model the volume fraction~$\phi^*$ is required to be infinitesimal. The advantage of this limitation is the possibility to derive the closed \gleich{eq:roscoeohnephic}.

\subsection{Assumptions}\label{sec:Assumptions}

In the following we assume that the particle phase consists of spheres with different diameters~$D_i$ that can be categorized in a finite number~$n$ of size or diameter classes. The size classes shall be sorted by diameter in ascending order, so that~$D_{i}<D_{i+1},~i=1,\dotsc,n-1$. The ratio of two consecutive diameters
\begin{equation}
\lambda_i\definiert\frac{D_{i+1}}{D_i}
\label{eq:uidef1}
\end{equation}
should be larger than~7 according to~\citet{1961_McGeary,2010_Brouwers} (or even~10 following~\citet{1971_Chong,2001_Dames}). In the completed suspension resulting from the construction process the $i$th size class occupies a volume~$V_i$ while the fluid phase occupies the volume~$V_f$. So the total volume~$\Vges$ of the suspension is given by
\begin{equation}
\Vges = V_f + \sum_{m=1}^n V_m
\label{eq:Vges}
\end{equation}
We assume that the suspension has an isotropic and homogeneous microstructure. As already outlined in \abschnitt{sec:basics}, the total volume fraction is defined by
\begin{equation}
\phi=\frac{V_{\text{particle}}}{V_{\text{fluid}}+V_{\text{particle}}}=\frac{\sum_{m=1}^n V_m}{\Vges}
\label{eq:phivol}
\end{equation}
Analogously it is useful to define volume fractions of single size classes, both during the construction process and in the completed suspension.

\subsection{Construction process}\label{sec:Aufbauprozess}

The models for apparent viscosity and maximum packing fraction that are developed in the following sections are based on the notion that the suspension is constructed by successive addition of new size classes. We call this process the \emph{construction process}. In the following, we focus on the volume fractions and generalize considerations in~\citet{2004_Pabst} and especially~\citet{1985_Norris}. Afterwards, we will use the expressions for the volume fractions to describe the relative viscosity. There are two possible approaches for the construction process:
\begin{description}
	\item[Variable total volume] In this case the volume of the fluid phase~$V_f$ is held constant during the construction process, so that the total volume of the suspension increases with each step until the suspension occupies the final volume~$\Vges$. The construction process thus only consists of additions of size classes.
	\item[Constant total volume] In order to keep the total volume of the suspension constant throughout the construction process, it is necessary to extract a suspension volume with a size equal to the added particle volume in each construction step (see also~\citet{1985_Norris}). So the extracted volume represents the composition of the existing suspension.\label{Verschmierung} 
\end{description}
Both approaches are equivalent as can be shown. In this work, we choose the case of constant total volume because of the intuitive meaning of volume fractions originating from the constant volume~$\Vges$. Considerations involving volumes may therefore easily be transferred to the notion of volume fractions, which is not true for the case of variable total volume.\par
The first addition of particles in the construction process implies a simple change in the total volume fraction of the particle phase~$\phi_i$:
\begin{equation}
\phi_0 = \frac{0}{\Vges}=0 \quad\longrightarrow \quad\phi_1 = \frac{V_1^*}{\Vges}
\label{eq:aufbaukonst1}
\end{equation}
$V_1^*$ denotes the added volume of the first particle size class which has to be distinguished from the volume of this class in the complete suspension according to the constant-volume approach. The more complex second construction step is given by
\begin{equation}
\phi_1 = \frac{V_1^*}{\Vges} \quad\longrightarrow\quad \phi_2 = \frac{\left(V_1^*-V_2^*\frac{V_1^*}{\Vges}\right)+V_2^*}{\Vges}
\label{eq:aufbaukonst2}
\end{equation}
In \gleich{eq:aufbaukonst2} the term in brackets represents the volume of the first size class still present after the second construction step. Therein the term~$V_2^*V_1^*/\Vges$ describes the loss of volume of the first size class due to the necessary extraction of volume. Before proceeding, we introduce the dimensionless notation
\begin{equation}
\phi_k^* = \frac{V_k^*}{\Vges}
\label{eq:phikstern}
\end{equation}
By rearranging of the expressions in \gleich{eq:aufbaukonst2} we find
\begin{align} \phi_1 = \underbrace{\phi_1^*}_{\definiertb\dphi_1^1}\quad\text{and} \label{eq:aufbaukonst2dim1}\\
\phi_2 = \underbrace{\phi_1^*(1-\phi_2^*)}_{\definiertb\dphi_1^2}+\underbrace{\phi_2^*}_{\definiertb\dphi_2^2}
\label{eq:aufbaukonst2dim2}
\end{align}
In Eqs.~\eqref{eq:aufbaukonst2dim1} and~\eqref{eq:aufbaukonst2dim2} we have introduced a notation for the volume fractions of the individual size classes during the construction process. The representation~$\dphi_k^i$ refers to the volume fraction of the~$k$th size class after the~$i$th construction step, that is after the addition of the~$i$th size class. So~$i$ means an index and no exponent. This should cause no confusion because the volume fraction will always occur linearly in all of the following expressions.\par
It can easily be shown, using Eqs.~\eqref{eq:aufbaukonst2dim1} and~\eqref{eq:aufbaukonst2dim2}, that the total volume fraction after the third construction step is given by
\begin{equation}\label{eq:phi3konstant}
\begin{split}
\phi_3 &= \phi_2(1-\phi_3^*)+\phi_3^* \\
			 &= \underbrace{\phi_1^*\klammera{2}\klammera{3}}_{\definiertb\dphi_1^3} + \underbrace{\phi_2^*\klammera{3}}_{\definiertb\dphi_2^3}+\underbrace{\phi_3^*}_{\definiertb\dphi_3^3}
\end{split}
\end{equation} 
The underlying pattern can be recognized clearly and so we may generalize intuitively:
\begin{equation}
\phi_{i+1} = \phi_i\klammera{i+1}+\phi_{i+1}^*
\label{eq:phiiplus1allg}
\end{equation}
Table \ref{tab:Aufbau} schematically outlines the construction process.\par
\renewcommand{\multirowsetup}{\centering}%
{\renewcommand{\baselinestretch}{1.0}
\begin{table*}[ht]
\centering
\tabcolsep0.5ex
\begin{tabular}{c c c c c c c}
\toprule
\multirow{2}{1.6cm}{After Step} & \multirow{2}{4cm}{Volume Fraction of Particle Phase} & \multirow{2}{*}{SC 1} & \multirow{2}{*}{SC 2} & \multirow{2}{*}{SC 3} & \multirow{2}{*}{$\dotso$} & \multirow{2}{*}{SC $n$} \\
 & & & & & & \\
\midrule
$ 1 $& $\phi_1 $&$ \dphi_1^1 $& & & & \\
\addlinespace
$2$ &  $\phi_2$ & $\dphi_1^2$& $\dphi_2^2$& & & \\
\addlinespace
$3$ &  $\phi_3$ & $\dphi_1^3$& $\dphi_2^3$& $\dphi_3^3$& & \\
$\vdots$ & $\vdots$& $\vdots$&$\vdots$ & $\vdots$ & $\ddots$& \\
$n$ & $\phi_n$ & $\dphi_1^n$& $\dphi_2^n$&$ \dphi_3^n$& $\dotso$ &$\dphi_n^n $\\ 
 \bottomrule    
\end{tabular}
\caption[Scheme for the volume fractions during the construction process]{Scheme for the volume fractions during the construction process (SC = size class)}
\label{tab:Aufbau}
\end{table*}
}
Up to now, the equations contain quantities determined by the construction process. However, in practical calculations one only knows the volume fractions~$\dphi_k^n$ of the size classes in the complete suspension. It is therefore useful to express the quantities describing the construction process by the composition of the complete suspension. For simplification, we define
\begin{equation}
\dphi_k \definiert \dphi_k^n
\label{eq:defdphi}
\end{equation}
These volume fractions in the complete suspension are given by
\begin{equation}
\dphi_k = \frac{V_k}{V_f+\sum_{m=1}^{n}V_m} = \frac{V_k}{\Vges}
\label{eq:dphiivol}
\end{equation} 
From Eqs.~\eqref{eq:aufbaukonst2dim1} to~\eqref{eq:phi3konstant} we deduce a general expression for~$\dphi_k^i$:
\begin{alignat}{2}
\dphi_k^i =& \phi_k^* \prod_{m=k+1}^i \klammera{m} &\quad\text{for } k<i \label{eq:dphikhochia}\\
\dphi_k^k =&  \phi_k^* &\quad\text{für } k=i \label{eq:dphikhochib}
\end{alignat}
We recall that there are no volume fractions with~$k>i$ (compare \tabelle{tab:Aufbau}). If we succeed in calculating the volume fractions~$\phi_k^*$, we are in a position to calculate the volume fractions of the individual size classes~$\dphi_k^i$ and the total volume fractions after each step~$\phi_i$ from Eqs.~\eqref{eq:dphikhochia},~\eqref{eq:dphikhochib} and from
\begin{equation}
\phi_i = \sum_{k=1}^i \dphi_k^i
\label{eq:summedphi}
\end{equation}
(consider \gleich{eq:phi3konstant}as an example). Writing down the volume fractions of the last size classes in the complete suspension and using definition~\eqref{eq:defdphi} as well as Eqs.~\eqref{eq:dphikhochia} and~\eqref{eq:dphikhochib} reveals the possibility to calculate the volume fractions~$\phi_k^*$:
\begin{align}
\dphi_n &= \phi_n^* \label{eq:dphin1}\\
\dphi_{n-1} &= \phi_{n-1}^*\klammera{n}\label{eq:dphin2}\\
\dphi_{n-2} &= \phi_{n-2}^*\klammera{n-1}\klammera{n}\label{eq:dphin3}\\
\dphi_{n-3} &=\dotso \nonumber
\end{align}
So we find by recursive insertion that
\begin{align}
\phi_n^* &= \dphi_n  \label{eq:phistern1}\\
\phi_{n-1}^* &= \frac{\dphi_{n-1}}{1-\dphi_n}\label{eq:phistern2}\\
\phi_{n-2}^* &= \frac{\dphi_{n-2}}{1-\dphi_n-\dphi_{n-1}}\label{eq:phistern3}\\
\phi_{n-3}^* &=\dotso \nonumber
\end{align}
This may be generalized in the form
\begin{equation}
\boxed{
\phi_k^* = \frac{\dphi_k}{1-\sum_{m=k+1}^n \dphi_m}}
\label{eq:phimstern}
\end{equation}
Inserting \gleich{eq:phimstern} into \gleich{eq:dphikhochia} yields
\begin{equation}
\boxed{
\dphi_k^i = \frac{\dphi_k}{1-\sum_{m=i+1}^n \dphi_m}}
\label{eq:dphikhochi3}
\end{equation}
So we have represented all of the quantities occurring in the construction process by the volume fractions in the complete suspension.

\subsection{A discrete model for the relative viscosity}\label{sec:Diskretes_Modell}

In this section, we apply the volume fraction relations derived above to the viscosity change during the construction process. As we have already noted, the differential \Bruggeman model lacks any information about the volume fraction of the individual particle size classes. For that reason we have described the construction process of the suspension in a discrete form in \abschnitt{sec:Aufbauprozess}. Preparing the development of the viscosity model, we first need to introduce the maximum packing fraction into the \Bruggeman model.\par
The \Roscoe equation~\eqref{eq:roscoeohnephic} diverges as the total volume fraction~$\phi$ approaches unity. In a real suspension, the achievable value of~$\phi$ is limited by the maximum packing fraction. In order to formally introduce the maximum packing fraction into the differential \Bruggeman model we proceed in a way proposed in~\citet{2009_Hsueh}. A similar way can be found in~\citet{2009_Mendoza}, where hard-sphere scaling is used (\citet{2003_Quin}). In both publications, it emerges that the notion of maximum packing fraction is introduced under little convincing considerations.\par
The approach followed in~\citet{2009_Hsueh} consists of modifying \gleich{eq:diffphistern} by using the maximum packing fraction~$\varphi_T$ in the description of the volume fraction of newly added particles~$\phi^*$. Therefore, it is supposed without derivation that
\begin{equation}
\phi^* = \frac{\dtotal\phi}{1-\frac{\phi}{\varphi_T}}
\label{eq:diffphisternphic}
\end{equation}
In combination with \gleich{eq:etaplusdeta} and after integration under the condition~$\eta_{\text{app}}(\phi=0) = \eta_0$ one finds the \Krieger relation~\eqref{eq:krieger1}
\begin{equation}
\eta_{\text{app}} = \eta_0 \left(1-\frac{\phi}{\varphi_T}\right)^{-\etaint{1}\varphi_T}
\label{eq:krieger}
\end{equation}
It would formally be possible to transfer the modification~\eqref{eq:diffphisternphic} to the discrete construction process, that is the volume fractions~\eqref{eq:phimstern}, and apply the result to the viscosity calculation. In the following it will be explained why this approach cannot be valid in general. Partially anticipating the later viscosity calculation, we raise two points.\par
Firstly, the construction process described in \abschnitt{sec:Aufbauprozess} is by no means dependent on the particle geometry. This is emphasized by the notion of homogenization between two construction steps. In contrast, the maximum packing fraction is strongly influenced by the particle geometry. So it would be artificial to introduce this quantity into the description of volume fractions during the construction process.\par
Secondly, the consideration of the volume fractions during the construction process is independent of the physical quantity that is calculated (here: the viscosity). It does not make any difference whether one calculates the viscosity or, for instance, the electric conductivity (or both at the same time). In both cases the construction process is constituted by the same volume fractions. Only through the employed relation between the volume fractions and the change in the physical quantity of interest parameters like the maximum packing fraction are included. This will be the approach followed during the later viscosity calculation.\par
The above considerations imply that the differential \Bruggeman model only allows for the derivation of the \Roscoe equation~\eqref{eq:roscoeohnephic} because as a consequence of this differential approach the right-hand side of \gleich{eq:etaplusdetaeingesetzt} may only consist of a linear expression (the \Einstein relation) that cannot contain the maximum packing fraction. So the approaches presented in~\citet{2009_Hsueh} and~\citet{2009_Mendoza} are formally possible but physically questionable. This disadvantage of the differential approach can be avoided in case of a discrete model, where finite volume fractions of particles are added during the construction process. The viscosity change is then described by nonlinear expressions containing the maximum packing fraction.\par
At this point it is necessary to introduce a distinct notation for the maximum packing fraction in order to avoid misinterpretations. The calculation of the maximum packing fraction will be conducted in \abschnitt{sec:Packungsdichte}. We choose the following notations partly referring to~\citet{2010_Brouwers}:\par
\noindent
\begin{tabularx}{\columnwidth}{lcX}
$\varphi_{Tk}^i$ & : & We denote as~$\varphi_{Tk}^i$ the maximum packing fraction of a polydisperse suspension consisting of~$k$ size classes after the~$i$th construction step ($i$th line in \tabelle{tab:Aufbau}). \\
$\varphi_T$ & : & The maximum packing fraction of the complete suspension is denoted as~$\varphi_T\definiert\varphi_{Tn}^n$. Hence, it is obvious why this notation has already been used in the correlations in \tabelle{tab:Korrelationen} and in the introduction.\\
$\varphi_c$ & : & We write~$\varphi_c\definiert\varphi_{T1}^i$ for the monodisperse packing fraction. The monodisperse maximum packing fraction~$\varphi_c$ is constant throughout the construction process and thus carries no upper index~$i$.\\
$\varphi_k^i$ & : & The volume fraction of the $k$th size class after the $i$th construction step in the state of maximum packing fraction is denoted by~$\varphi_k^i$.\\
\end{tabularx}\par
\noindent This notation is visualized in \tabelle{tab:Packung}(compare \tabelle{tab:Aufbau}).
\renewcommand{\multirowsetup}{\centering}%

{\renewcommand{\baselinestretch}{1.0}
\begin{table*}[ht]
\centering
\tabcolsep0.5ex
\begin{tabular}{c c c c c c c c c c c}
\toprule
\multirow{2}{1.6cm}{After Step} & \multirow{2}{4cm}{Volume Fraction of Particle Phase} & \multirow{2}{*}{SC 1} & & \multirow{2}{*}{SC 2} & & \multirow{2}{*}{SC 3} & & \multirow{2}{*}{$\dotso$} & & \multirow{2}{*}{SC $n$} \\
 & & & & & & & & & & \\
\midrule
$1$ &  $\phi_1 $& $ \varphi_c$ & & & & & & & & \\
\addlinespace
$2$ &  $\phi_2$ & $\varphi_c$& $\to$ & $\varphi_{T2}^2$& & & & & & \\
\addlinespace
$3$ &  $\phi_3$ & $\varphi_c$& $\to$ & $\varphi_{T2}^3$& $\to$ & $\varphi_{T3}^3$& & & &\\
$\vdots$ & $\vdots$& $\vdots$ & & $\vdots$& & $\vdots$& & &  & \\
$n$ & $\phi_n$ & $\varphi_c$& $\to$ & $\varphi_{T2}^n$& $\to$ & $ \varphi_{T3}^n$& $\to$ & $\dotso$ & $\to$ & $\varphi_{Tn}^n $\\ 
 \bottomrule    
\end{tabular}
\caption[Notation for the maximum packing fraction during the construction process]{Notation for the maximum packing fraction during the construction process (SC = size class); the arrows indicate that the far right values are calculated recursively from the previous values (for details of the calculation see \abschnitt{sec:Packungsdichte})}
\label{tab:Packung}
\end{table*}
}
After each construction step the maximum packing fraction is newly calculated according to the new composition of the suspension. This is conducted within a recursive process (arrows in \tabelle{tab:Packung}). Starting from the monodisperse maximum packing fraction~$\varphi_c$ all previously added size classes are taken into account and so for each step~$i$ the value~$\varphi_{Ti}^{i}$ is calculated. This value is needed in \gleich{eq:allgvisko}, which is still to derive.\par
We now build on the description of the discrete construction process given in \abschnitt{sec:Aufbauprozess} deriving the viscosity relations. When during the construction process the $(i+1)$th size class is added, the total volume fraction of the particle phase changes from~$\phi_i$ to~$\phi_{i+1}$. This is associated with a change in apparent viscosity from~$\eta_i$ to~$\eta_{i+1}$. To emphasize the analogy to the differential \Bruggeman model, we temporarily confine ourselves to the linear \Einstein relation as a description of the viscosity change and afterwards we will extend the model to higher orders.\par
According to \abschnitt{sec:Aufbauprozess} the volume fraction of the newly added $(i+1)$th size class is denoted as $\phi_{i+1}^*$, so the new apparent viscosity is given by
\begin{equation}
\eta_{i+1} = \eta_i \left( 1+\etaint{1} \phi_{i+1}^* \right)
\label{eq:etasterniplus1}
\end{equation}
The explicit form of \gleich{eq:etasterniplus1} is
\begin{equation}
\eta_{i+1} = \eta_0\prod_{m=1}^{i+1}  \left( 1+\etaint{1}\phi_{m}^* \right)
\label{eq:expetasternmitdphi}
\end{equation}
For $i+1 = n$ we find the apparent viscosity~\mbox{$\eta=\eta_n$} of the complete suspension (it is well known that $\sum_{l=n+1}^{n} x = 0$ for all $x$).\par
The differential \gleich{eq:etaplusdetaeingesetzt} contains only the first-order \Einstein relation because of the infinitesimal character of~$\dtotal\phi$. In the case of the discrete model represented by \gleich{eq:expetasternmitdphi} it is not necessary to confine oneself to linear terms. Therefore, we are allowed to describe the modification of the apparent viscosity more accurately by higher order terms of~$\phi_{k}^*$. So \gleich{eq:expetasternmitdphi} can be extended according to
\begin{equation}
\boxed{
\eta_{n} = \eta_0\prod_{m=1}^{n}  \left[ 1+\etaint{1}\phi_{m}^* +\etaint{2}\left(\phi_{m}^*\right)^2+ \dotsb\right]\label{eq:allgetasterniplus1}}
\end{equation}
In this way, one can establish a connection between \gleich{eq:allgetasterniplus1} and models existing in the literature. Selecting the coefficients from the \Taylor expansion of a viscosity relation (\eqref{eq:viskogleichung}, for instance) at the point~$\phi=0$ for the coefficients~$\etaint{k}$ in \gleich{eq:allgetasterniplus1} and formally considering an infinite number of terms, we may substitute the bracketed term in \gleich{eq:allgetasterniplus1} with the viscosity relation itself. This step requires the introduction of the maximum packing fraction. Using a viscosity relation in the sense of \gleich{eq:viskogleichung} we find
\begin{equation}
\boxed{
\eta_{n}=\eta_0\prod_{m=1}^{n}\eta_r(\phi_{Pm},\varphi_{Tm}^{m})
\label{eq:allgvisko}}
\end{equation}
In \gleich{eq:allgvisko}~$\varphi_{Tm}^{m}$ refers to the maximum packing fraction after the~$m$th construction step that will be calculated in \abschnitt{sec:Packungsdichte}. Writing down \gleich{eq:allgvisko} we make the important assumption that the viscosity change in each step~$m$ depends on the actual maximum packing fraction value~$\varphi_{Tm}^{m}$ and not only on the final value~$\varphi_T=\varphi_{Tn}^{n}$.\par
Despite the use of the variable maximum packing fraction \gleich{eq:allgvisko} corresponds to the so-called \Farris model (\citet{1968_Farris}) also referred to in~\citet{2010_Brouwers,1993a_Sudduth,1990_Cheng}. So we have found an interesting connection between the differential \Bruggeman model and the \Farris model which is drawn by the discrete construction process.\par
It is important to note that in extending \gleich{eq:allgetasterniplus1} to~\eqref{eq:allgvisko} we have considered the complete \Taylor expansion of relation~\eqref{eq:viskogleichung}, suggesting the admissibility of arbitrary volume fractions of the individual size classes which is uncertain regarding the notion of the construction process. Since it is not possible to state a limiting value of~$\phi_{m}^*$ for the validity of the result~\eqref{eq:allgvisko}, we will use this equation at least as a reasonable approximation also for higher values of~$\phi_{m}^*$.

\subsection{Determination of the maximum packing fraction}\label{sec:Packungsdichte}

In the following we review two models for computing the polydisperse maximum packing fraction reported in the literature. The first model has been proposed by \citet{1931_Furnas} (see also~\citet{2010_Brouwers,1993b_Sudduth}) and the second one, using \Furnas' results, by \citet{1993b_Sudduth}. We then derive a model which modifies the \Furnas approach and serves as an alternative to the \Sudduth model for large diameter ratios. During the calculation we will use the quantities occurring in the construction process according to Tab.~\ref{tab:Aufbau} and~\ref{tab:Packung}. The new model for the maximum packing fraction can be easily introduced into the viscosity model (as explained in the context of \gleich{eq:allgvisko}) because it is conceptually consistent with the construction process. This is an advantage over models that use a maximum packing fraction that has been derived or measured entirely independent of the viscosity relation in which it is to be used (e.g.~\citet{2001_Dames,1971_Chong} or~\citet{1988_Poslinski} using a model taken from~\citet{1984_Ouchiyama}). The same applies for the model proposed by \citet{2009_Farr}, which is based upon a mapping of the 3D packing problem onto a 1D problem allowing for inexpensive estimation of the maximum packing fraction. This model is very promising even for finite size ratios but still needs a non-trivial sorting-like algorithm. However, a close look at that algorithm reveals that it is very similar to the construction process used in the present work and therefore offers the possibility to be used as an alternative, especially when it comes to small size ratios.

\subsubsection{The Furnas model}\label{sec:DasModellvonFurnas}

The \Furnas model estimates the highest possible value of the maximum packing fraction for a given number of size classes with large diameter ratios (cf. \abschnitt{sec:Assumptions}). It is based on the assumption that the state of maximum packing fraction is constructed successively from size classes with decreasing diameter. At first, the larger spheres fill the entire available suspension volume with the monodisperse maximum packing fraction~$\varphi_c$. So the total volume fraction is
\begin{equation}
\varphi_{T1,\Furnas}^i = \varphi_1^i = \varphi_c
\label{eq:FurnasT0}
\end{equation}
Subsequently, the remaining volume fraction~$(1-\varphi_c)$ is filled by the spheres with the next smaller diameter, also to a realizable part of~$\varphi_c$. So the additional volume fraction of small spheres~$\varphi_2^i$ is given by
\begin{equation}
\varphi_{T2}^i = \varphi_1^i+\varphi_2^i = \varphi_c + (1-\varphi_c)\varphi_c = 1 - (1-\varphi_c)^2
\label{eq:FurnasT1}
\end{equation}
Generalizing these considerations to a number of~$n$ size classes, one finds (\citet{1993b_Sudduth})
\begin{equation}
\varphi_{Tn,\Furnas}^n = 1 - (1-\varphi_c)^{n}
\label{eq:FurnasTn}
\end{equation}
In the \Furnas model the maximum packing fraction is calculated under the assumption that the size distribution allows that each size class fills the entire interstitial space to a fraction of~$\varphi_c$. This corresponds to a certain size distribution for a given number of size classes~\citet{1961_McGeary}.

\subsubsection{The Sudduth model}

To account explicitly for the size distribution, \citet{1993b_Sudduth} proceeded as follows. Starting from experimental data by~\citet{1961_McGeary} for bidisperse suspensions, \Sudduth found that the maximum packing fraction of polydisperse suspensions can be parameterized using certain moments~($\mathcal{D}_x$ for the $x$th moment) of the particle size distribution
\begin{equation}
\mathcal{D}_x=\frac{\sum_{i=1}^{n} N_i D_i^x}{\sum_{i=1}^n N_i D_i^{x-1}}
\label{eq:Dx}
\end{equation}
where~$D_i$, as before, is the particle diameter and~$N_i$ the number density of the~$i$th size class, respectively. In conjunction with the maximum packing fraction value~\eqref{eq:FurnasTn} from the \Furnas model, \Sudduth ended up with the approximation
\begin{multline}
\varphi_{Tn,\Sudduth}^n=\varphi_{Tn,\Furnas}^n\\-\left(\varphi_{Tn,\Furnas}^n-\varphi_c\right)\,\euler{0.27\left(1-\mathcal{D}_5/\mathcal{D}_1\right)}
\label{eq:Sudduth}
\end{multline}
for the complete suspension, where~$\mathcal{D}_5,~\mathcal{D}_1$ are given by \gleich{eq:Dx}. We will compare the \Sudduth model to our results in \abschnitt{sec:Vergleich}. Note that, despite the model~\eqref{eq:Sudduth} can possibly cover all kinds of size distribution, it lacks a theoretical foundation. Therefore, we attempt to propose an alternative model for large size ratios consistent with the construction process and the viscosity model~\eqref{eq:allgvisko}.

\subsubsection{A new model for the maximum packing fraction}\label{sec:DasneueModell}

We will derive the new model in a way that first the instructive case of bidisperse suspensions is treated explicitly in order to prepare the subsequent general derivation. It should be noted that the bidisperse model presented below has already been proposed by several authors (for instance~\citet{1987_Yu,1997_Gondret}), but it is helpful to discuss it in detail to prepare the polydisperse derivation (which appears to be novel).\par
A consequent calculation of the maximum packing fraction can be achieved by retaining the particle size distribution existing in the fluent suspension state during the composition of the maximum packing fraction. This main assumption can be expressed in the form
\begin{equation}
\frac{\varphi_{k+1}^i}{\varphi_k^i} = \frac{\dphi^i_{k+1}}{\dphi^i_k}
\label{eq:verhaeltnis}
\end{equation}
where~$\varphi_k^i$ in the state of maximum packing fraction corresponds to~$\dphi^i_k$ in the fluent state, that is the volume fraction of the~$k$th size class.
\setcounter{paragraph}{0}
\subparagraph{Derivation for bidisperse systems}

Bidisperse (or bimodal) suspensions contain two different particle sizes. Given the ratio
\begin{equation}
\frac{\varphi_2^i}{\varphi_1^i} = \frac{\dphi^i_{2}}{\dphi^i_1}
\label{eq:alpha1a}
\end{equation}
we ask how the state of maximum packing fraction can be achieved retaining the volume fraction ratio~\eqref{eq:alpha1a}. There are two possibilities differing with respect to the value of~$\dphi^i_2$. Both situations are visualized in \bild{fig:Bimodal}.
\begin{figure}[ht]%
\centering%
\subfigure[]{\label{fig:Bimodal1}
\includegraphics[width=4cm]{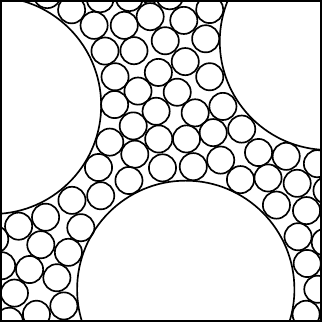}
}\hspace{1.5cm}
\subfigure[]{\label{fig:Bimodal2}
\includegraphics[width=4cm]{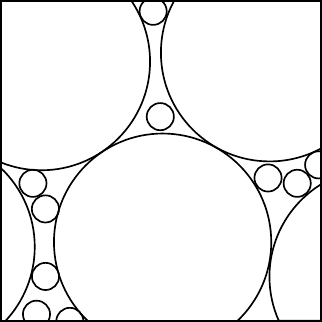}
}
\caption[Schematic representation of the two possible situations for the bidisperse maximum packing fraction]{Schematic representation of the two possible situations for the bidisperse maximum packing fraction: \protect\subref{fig:Bimodal1}~Situation~\eins: The fraction of large particles lies below~$\varphi_c$ while the small particles fill the interstices with a fraction of~$\varphi_c$. \protect\subref{fig:Bimodal2}~Situation~\zwei: The large particles occupy a fraction of~$\varphi_c$ while the small particles are present in the interstices according to the volume fraction ratio.}%
\label{fig:Bimodal}%
\end{figure}

\paragraph{Situation~\eins}

The volume fraction~$\varphi_2^i$ of the large spheres lies below the monodisperse packing fraction~$\varphi_c$ and the remaining interstice~$1-\varphi_2^i$ is filled to an amount of~$\varphi_c$ with small spheres, compare \bild{fig:Bimodal1}. So the total particle volume fraction is
\begin{equation}
\varphi_{T2}^{i\eins} = \varphi_1^i+\varphi_2^i = \varphi_2^i + (1-\varphi_2^i)\varphi_c
\label{eq:IchT2}
\end{equation}
(compare \gleich{eq:FurnasT1}, where restrictively~$\varphi_2^i = \varphi_c$). Using \gleich{eq:alpha1a} we find from \gleich{eq:IchT2}
\begin{equation}
\varphi_2^i = \frac{\varphi_c}{\varphi_c + \frac{\dphi_1^i}{\dphi^i_2}}
\label{eq:phi2bi}
\end{equation}
and furthermore
\begin{equation}\label{eq:phiT1sit1}
\varphi_{T2}^{i\eins} = \frac{\varphi_c(\dphi_2^i+\dphi_2^i)}{\varphi_c\dphi^i_2 + \dphi_1^i} 
\end{equation}
The limitation for the validity of \gleich{eq:phiT1sit1} follows from the realizability condition ~\mbox{$\varphi_2^i<\varphi_c$} (arbitrarily, we assign the equal sign to the situation~\zwei in order to avoid an ambiguous definition of the case~$\varphi_2^i=\varphi_c$\label{fuss}) which in combination with \gleich{eq:phi2bi} yields
\begin{equation}
\dphi^i_2 < \frac{\dphi_1^i}{1-\varphi_c}
\label{eq:alphagrenz1sit1}
\end{equation}
as the condition for the validity of situation~\eins.

\paragraph{Situation~\zwei}

In this situation the volume fraction~$\varphi_2^i$ of the large spheres equals the monodisperse maximum packing fraction~$\varphi_c$ (it is not possible to exceed the value of~$\varphi_c$ in a monodisperse loading) and so we have~$\varphi_2^i = \varphi_c$, see \bild{fig:Bimodal2}. Using \gleich{eq:alpha1a} we immediately find
\begin{align} \varphi_1^i&=\varphi_2^i\frac{\dphi_1^i}{\dphi^i_2}\quad\text{and} \label{eq:phi1sit2}\\
\varphi_{T2}^{i\zwei} &= \varphi_1^i+\varphi_2^i = \varphi_c\left(1+\frac{\dphi_1^i}{\dphi^i_2}\right)
\label{eq:phiT1sit2}
\end{align}
It follows from the fundamental requirement~$\varphi_{T2}^{i\zwei}<1$ and \gleich{eq:phiT1sit2} that
\begin{equation}
\dphi^i_2>\frac{\varphi_c\dphi_1^i}{1-\varphi_c}
\label{eq:alpha1schwach}
\end{equation}
whereas the realizability condition~$\varphi_1^i\leq(1-\varphi_2^i)\varphi_c$ under consideration of~$\varphi_2^i=\varphi_c$ and \gleich{eq:phi1sit2} yields
\begin{equation}
\dphi^i_2\geq\frac{\dphi_1^i}{1-\varphi_c}
\label{eq:alphagrenz1sit2}
\end{equation}
The condition \eqref{eq:alphagrenz1sit2} includes the \ungleich{eq:alpha1schwach}. So \gleich{eq:alphagrenz1sit2} serves as a condition for the validity of situation~\zwei.

\subparagraph{Generalization for polydisperse systems}
\setcounter{paragraph}{0}
In the following, we will generalize the model for polydisperse systems which corresponds to deriving expressions for the total volume fraction~$\varphi_{Tk+1}^i$ in the situations~\eins and~\zwei and for the limiting value of~$\dphi^i_{k+1}$. Regardless of the number of size classes, there are always two situations. This relies on the geometric fact that no particle class except the one with the largest particle diameter can reach the monodisperse packing fraction. If hypothetically a particle class with a smaller diameter reached the monodisperse maximum packing fraction, the larger particles would not fit into the interstices between the smaller particles and could thus not contribute to this state of maximum packing fraction.

\paragraph{Situation~\eins}

In the polydisperse case the interstices between the largest particles are filled by the smaller particle size classes with the packing fraction~$\varphi_{Ti}^k$ which is determined by the volume fractions~$\dphi^i_1$ to~$\dphi^i_k$. So the volume fraction of the largest particles is~$\varphi_{k+1}^i$ and the one of all the smaller particles is~$(1-\varphi_{k+1}^i)\varphi_{Tk}^i$. The sum of both fractions yields the maximum packing fraction, that is
\begin{equation}
\varphi_{Tk+1}^{i\eins} = \varphi_{k+1}^i + (1-\varphi_{k+1}^i)\varphi_{Tk}^i
\label{eq:phiTiplus1urspr}
\end{equation}
So we may express the volume fractions of all the small particles by
\begin{equation}
\sum_{m=1}^k \varphi_m^i = (1-\varphi_{k+1}^i)\varphi_{Tk}^i
\label{eq:summephik}
\end{equation}
Eq.~\eqref{eq:verhaeltnis} allows for representing all the occurring volume fractions by the fraction~$\varphi_{k+1}^i$ of the largest particles, so we can reformulate \gleich{eq:summephik}:
\begin{equation}
\sum_{m=1}^k \varphi_m^i = \varphi_{k+1}^i \einsdurch{\dphi_{k+1}^i} \sum_{m=1}^k \dphi^i_m = (1-\varphi_{k+1}^i)\varphi_{Tk}^i
\label{eq:summephik2}
\end{equation}
Solving \gleich{eq:summephik2} for $\varphi_{k+1}^{i\eins}$ and inserting the result into \gleich{eq:phiTiplus1urspr} yields the maximum packing fraction in situation~\eins:
\begin{equation}
\boxed{
\varphi_{Tk+1}^{i\eins} = \frac{\varphi_{Tk}^i\sum_{m=1}^{k+1}\dphi_m^i}{\varphi_{Tk}^i\dphi^i_{k+1} + \sum_{m=1}^{k}\dphi_m^i}
\label{eq:phiTiplus11}
}
\end{equation}
The limiting value for~$\dphi^i_{k+1}$ follows analogously to the bi- and tridisperse cases from the requirement~$\varphi_{k+1}^i<\varphi_c$ and is thus given by
\begin{equation}
\dphi^i_{k+1} < \frac{\varphi_c\sum_{m=1}^{k}\dphi_m^i}{\varphi_{Tk}^i(1-\varphi_c)}
\label{eq:alphagrenzsit1}
\end{equation}

\paragraph{Situation~\zwei}

In situation~\zwei the largest particles occupy a volume fraction equal to the monodisperse maximum packing fraction, so~$\varphi_{k+1}^i=\varphi_c$. Therefore, by means of \gleich{eq:alpha1a} we may write
\begin{equation}
\boxed{
\varphi_{Tk+1}^{i\zwei} = \varphi_c \frac{\sum_{m=1}^{k+1}\dphi_m^i}{\dphi^i_{k+1}}
\label{eq:phiTiplus12}
}
\end{equation}
for the maximum packing fraction in situation~\zwei. The limiting value of~$\dphi^i_{k+1}$ in this situation is determined by the condition
\begin{equation}
\sum_{m=1}^k \varphi_m^i \leq (1-\varphi_c)\varphi_{Tk}^i
\label{eq:forderung2}
\end{equation}
and combined with~\eqref{eq:summephik} and~\eqref{eq:phiTiplus12} takes the form
\begin{equation}
\dphi^i_{k+1} \geq \frac{\varphi_c\sum_{m=1}^{k}\dphi_m^i}{\varphi_{Tk}^i(1-\varphi_c)}
\label{eq:alphagrenzsit2}
\end{equation}
\paragraph{Properties of the bounds}\label{Schranken}

The bounds~\eqref{eq:alphagrenzsit1} and~\eqref{eq:alphagrenzsit2} can alternatively be derived by equaling the prescriptions~\eqref{eq:phiTiplus11} and~\eqref{eq:phiTiplus12}. This underlines the consistency between the situations~\eins and~\zwei because there is a continuous transition. This is equivalent to choosing the smallest value out of the two calculated maximum packing fractions because for all values of~$\dphi_{k+1}^i$ the situation that is present is always the one with the smaller maximum packing fraction. Thus we are lead to the prescription
\begin{equation}
\boxed{
\varphi_{Tk+1}^i=\text{min}\left[\varphi_{Tk+1}^{i\eins},\varphi_{Tk+1}^{i\zwei}\right]
}
\label{eq:minphiT1}
\end{equation}
%
%%#############################################################################################
%
\subsection{Comparison with experiments and models from the literature}\label{sec:Vergleich}

In this section, we compare our model to experimental data and models reported in the literature. \citet{1961_McGeary} conducted experiments to determine the maximum packing fraction of polydisperse compositions of spheres. To achieve high packing fractions, he investigated systems of up to four particle size classes mostly with a large diameter ratio (cf.~\abschnitt{sec:Assumptions}). Note that we consider his data without applying the manipulation used by~\citet{1993b_Sudduth}, which would not change the result of our comparison. \bild{fig:1961_McGeary_3} compares the bidisperse maximum packing fraction calculated from Eqs.~\eqref{eq:phiT1sit1} and~\eqref{eq:phiT1sit2}, depending on the criteria~\eqref{eq:alphagrenz1sit1} and \eqref{eq:alphagrenz1sit2}, respectively, with experimental data for three different large diameter ratios. 
\begin{figure}[ht]%
\centering%
\includegraphics{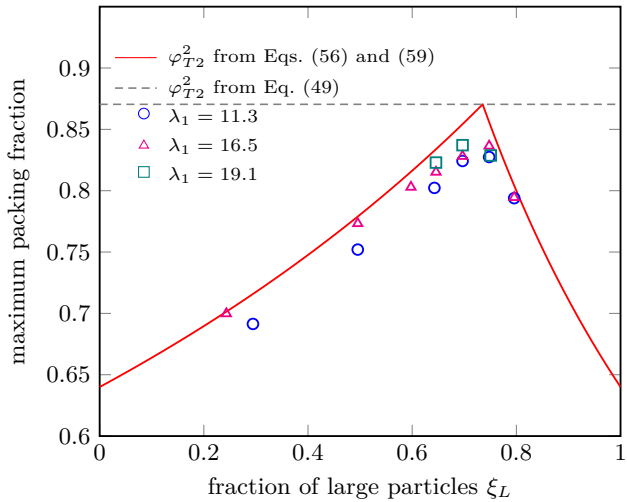}%
\caption{Comparison of Eqs.~\eqref{eq:phiT1sit1} and~\eqref{eq:phiT1sit2} for the maximum packing fraction of bidisperse suspension with the data from~\citet{1961_McGeary} using~$\varphi_c=0.64$, accordingly; the \Furnas model~\eqref{eq:FurnasT1} gives the maximum value of~$\varphi_{T2,\Furnas}^2=0.8704$; $\lambda_1$ is the diameter ratio~\eqref{eq:uidef1}}%
\label{fig:1961_McGeary_3}%
\end{figure}
All experimental values lie below the curve given by the model, indicating that the latter correctly predicts the bidisperse large size ratio limit of the maximum packing fraction. In addition, we calculate the maximum packing fraction for the quaternary system from~\citet{1961_McGeary}, as well as for the corresponding bi- and tridisperse composition, and compare it to the experimental values in \tabelle{tab:1961_McGeary}.
\begin{table*}[ht]%
\centering%
\begin{tabular}{c@{\extracolsep{-0.03ex}}c@{\extracolsep{-0.5ex}}ccccc@{\extracolsep{-0.5ex}}c@{\extracolsep{-0.5ex}}c@{\extracolsep{-0.5ex}}c@{\extracolsep{-1ex}}}
\toprule
\multirow{2}{*}{$i$} & \multirow{2}{*}{$\lambda$} & \multirow{2}{*}{$\displaystyle\frac{\dphi_1}{\phi}$} & \multirow{2}{*}{$\displaystyle\frac{\dphi_2}{\phi}$} & \multirow{2}{*}{$\displaystyle\frac{\dphi_3}{\phi}$} & \multirow{2}{*}{$\displaystyle\frac{\dphi_4}{\phi}$} & \multirow{2}{2.1cm}{\centering $\varphi_{Tn}^n$\newline \McGeary} & \multirow{2}{2.1cm}{\centering $\varphi_{Tn}^n$ \newline this work}  & \multirow{2}{2.1cm}{\centering $\varphi_{Tn}^n$ \Sudduth} & \multirow{2}{2.1cm}{\centering $\varphi_{Tn}^n$ \Furnas} \\
 & & & & & & & & & \\
\toprule
4 & \multirow{4}{0.8cm}{8.3\newline5.4\newline7} &  &  &  & 1.000 & 0.580 & 0.580 & 0.580 & 0.580\\
3 &  &       &       & 0.274 & 0.726 & 0.800 & 0.799 & 0.784 & 0.824\\
2 &  &       & 0.109 & 0.244 & 0.647 & 0.898 & 0.896 & 0.926 & 0.926\\
1 &  & 0.061 & 0.102 & 0.230 & 0.607 & 0.951 & 0.956 & 0.969 & 0.969\\
\bottomrule
\end{tabular}
\caption{Maximum packing fraction~$\varphi_{Tn}~(n=1,2,3,4)$ from~\citet{1961_McGeary} compared against three different models: This work (Eqs.~\eqref{eq:phiTiplus11} and~\eqref{eq:phiTiplus12}, respectively), the model by \citet{1993b_Sudduth}, and the \Furnas model (Eq.~\eqref{eq:FurnasTn}); $i$~denotes the index of a size class, $\lambda$~is the diameter ratio between two consecutive classes}
\label{tab:1961_McGeary}
\end{table*}
Therein, also the limiting values provided by the \Furnas model (\abschnitt{sec:DasModellvonFurnas}) are given. The fact that our model represents the experimental data more accurately than the \Furnas model shows that the theoretical limit does not coincide with \McGeary's densest packing but lies very close to it. Also the model by~\citet{1993b_Sudduth} deviates more strongly from the experimental data than our model. Within the experimental uncertainties of the data, this shows that our model for the maximum packing fraction is at least equivalent to the \Sudduth model for large size ratios. Note that the smallest size ratio of~$\lambda=5.4$ dies not fulfill the condition from~\abschnitt{sec:Assumptions}. However, this has a negligible effect in the considered case because the two corresponding size classes (2 and~3 in \tabelle{tab:1961_McGeary}) together occupy less than one third of the total solid volume fraction.\par
For comparison of the complete viscosity model with experimental data we choose the case of bidisperse suspensions. In~\citet{1988_Poslinski}, experimental data from~\citet{1971_Chong,1959_Sweeny} as well as original data for the relative viscosity of bidisperse suspensions with large size ratios ($\lambda_1=5.2,~7.2,~20.8$) for several total volume fractions is provided.%
In order to apply the discrete viscosity model as presented above to the experimental data, we proceed as follows (cf. \bild{fig:Ablauf}). First, we determine the parameters of the underlying monodisperse viscosity relation, for which we choose \gleich{eq:viskogleichung} (this choice is due to consistency with~\citet{1988_Poslinski}, where the \Quemada relation~\eqref{eq:Quemada} is used, alternative choices are given in \tabelle{tab:Korrelationen}). We set~$\etaint{1}=2.5$, $\etaint{2}=5.2$ (according to~\citet{1972_Batchelor}) and~$\varphi_c=0.64$. Second, we evaluate the model Eqs.~\eqref{eq:allgvisko},~\eqref{eq:phiT1sit1} and~\eqref{eq:phiT1sit2} for bidisperse systems ($n=2$), thereby introducing the fraction of large particles
\begin{equation}
\xi_L\definiert\frac{\dphi_2^2}{\dphi_1^2+\dphi_2^2}\definiertb\frac{\dphi_2^2}{\phi}
\label{eq:xiL}
\end{equation}
with the total volume fraction~$\phi$. The viscosity \gleich{eq:allgvisko} can thus be written as
\begin{multline}
\eta_{2} = \eta_0\prod_{m=1}^{2}\left[\left(\etaint{1}-\frac{2}{\varphi_{Tm}^{m}}\right)\phi_{m}^*\right.\\+\left.\left(\etaint{2}-\frac{3}{(\varphi_{Tm}^{m})^2}\right)(\phi_{m}^*)^2+\left(1-\frac{\phi_{m}^*}{\varphi_{Tm}^{m}}\right)^{-2}\right]
\label{eq:bimod1}
\end{multline}
where $\phi_{1}^*=\dphi_1/(1-\dphi_2)=\phi\xi_L/(1-\phi\xi_L)$ and $\phi_{2}^*=\dphi_1=\phi\xi_L$ are calculated from \gleich{eq:phimstern} (recall that~$\dphi_1^2=\dphi_1$ and~$\dphi_2^2=\dphi_2$ for bidisperse systems, cf. \gleich{eq:dphikhochi3}). The maximum packing fraction in each of the situations~\eins and~\zwei (see \abschnitt{sec:Packungsdichte}) according to Eqs.~\eqref{eq:phiT1sit1} and~\eqref{eq:phiT1sit2}, respectively, may be written as
\begin{align}
\varphi_{T2}^{2\eins} & =\frac{\varphi_c\phi}{\varphi_c\dphi_2+\dphi_1} =\frac{\varphi_c}{\varphi_c\xi_L+1-\xi_L}\\
\varphi_{T2}^{2\zwei} & =\varphi_c\frac{\phi}{\dphi_2} =\frac{\varphi_c}{\xi_L}
\label{eq:bimod2}
\end{align}
Situations~\eins or~\zwei are chosen from condition~\eqref{eq:minphiT1}
\begin{equation}
\boxed{
\varphi_{T2}^2=\text{min}\left[\varphi_{T2}^{2\eins},\varphi_{T2}^{2\zwei}\right]
}
\label{eq:bimod3}
\end{equation}
\bild{fig:1988_Poslinski_3} shows good agreement between theory and experiment especially for the highest volume fractions~$\phi$, but an underestimation of the effect of bidispersity, that is the viscosity minimum, for the two smallest volume fractions.
\begin{figure}[ht]%
\centering%
\includegraphics{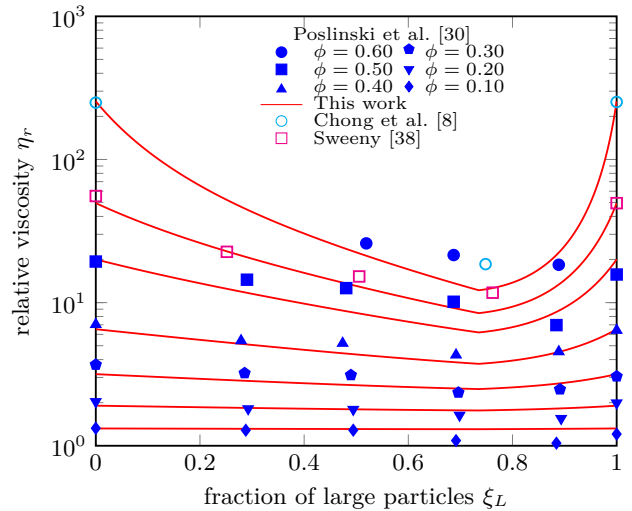}%
\caption{Comparison of the model consisting of Eqs.~\eqref{eq:allgvisko} for~$n=2$,~\eqref{eq:phiT1sit1} and~\eqref{eq:phiT1sit2} with experimental data from~\citet{1988_Poslinski,1971_Chong,1959_Sweeny} for various total volume fractions~$\phi$ ($\etaint{1}=2.5,~\etaint{2}=5.2,~\varphi_c=0.64$); missing total volume fractions are~$\phi=0.60$ (\citet{1971_Chong}) and~$\phi=0.55$ (\citet{1959_Sweeny}); diameter ratios are $\lambda_1=5.2$ (\citet{1988_Poslinski}), $\lambda_1=7.2$ (\citet{1971_Chong}) and $\lambda_1=20.8$ (\citet{1959_Sweeny}), respectively}%
\label{fig:1988_Poslinski_3}%
\end{figure}
Note that, despite the small size ratio of~$\lambda_1=5.2$ in \Poslinski's data, our model can be applied as a good approximation. We may conclude that the model presented in this work correctly displays the behavior of bimodal suspensions with large size ratio, which implies that the way in which the maximum packing fraction has been introduced into the construction process is reasonable. It must be emphasized that the viscosity reduction with respect to a monodisperse suspension, as shown in \bild{fig:1988_Poslinski_3} by our model, is not only due to the increased maximum packing fraction for $0<\xi<1$, but also due to the interaction of the two particle size classes by means of the excluded volume, that is, the volume fractions $\phi_{m}^*$ in \gleich{eq:bimod1}.  
\section{Conclusions}\label{chap:conclusions}

In the present work, we proposed a model for the relative viscosity of polydisperse suspensions of non-colloidal hard spheres. Using monodisperse viscosity correlations, we described polydisperse suspensions by means of a construction process consisting of successive additions of particle size classes.\par
As a starting point, we proposed generalized forms of the well-known \Quemada and \Krieger equations that allow for the choice of the second order intrinsic viscosity~$\etaint{2}$. These modified equations are more suitable to describe the relative viscosity over the whole range of concentrations than the original relations.\par
Later, we described the construction process in detail applying a dimensionless way of description based on volume fractions. This rigorous description served as a basis for the calculation of the relative viscosity during the construction process. Starting from the \Bruggeman model, we finally arrived at the \Farris model, connecting two approaches commonly regarded as uncorrelated.\par
As an entirely new component, we introduced the polydisperse maximum packing fraction into the \Farris model. The way of introducing the maximum packing fraction into the effective medium approach has been justified by conceptual considerations regarding the construction process. Consistently with the relative viscosity calculation and therefore in contrast to most approaches in the literature, we derived a formalism to determine the polydisperse maximum packing fraction by means of the same construction process. It is assumed that the maximum packing fraction affects the viscosity in each step of the construction process.\par
Comparing separately the maximum packing fraction model and the complete viscosity model to experimental data from the literature, good agreement has been achieved. We showed that for hard-sphere suspensions with large diameter ratios the empirical \Sudduth model can be replaced by a maximum packing fraction model based on an intuitive geometrical argumentation. This could offer the possibility to modify the \Sudduth model using the model presented here as a starting point.\par
Additionally, we revealed a possible approach for integrating particle deformability, represented by a particle phase viscosity, into the viscosity model using a result from the literature.\par
So far, our model is only valid for large diameter ratios of consecutive size classes during the construction process. An attempt to generalize the present model, which fully accounts for the size distribution in the limiting case of large size ratios, to the case of small size ratios will be presented in a future work.
%\begin{appendix}
%%
%%\appendix
%\section{Scheme for the viscosity estimation}\label{app:A}
%%
%
%
%\end{appendix}

\bibliographystyle{plainnat}

\end{document}